\documentclass[universe,article,accept,moreauthors,pdftex,10pt,a4paper]{mdpi} 
\firstpage{1} 
\articlenumber{32}
\doinum{10.3390/universe2040032}
\pubvolume{4}
\pubyear{2016}
\copyrightyear{2016}
\externaleditor{Academic Editors: Roman Pasechnik, José Eliel Camargo-Molina and António Pestana Morais}
\history{Received: 14 October 2016; Accepted: 1 December 2016; Published: 13 December 2016}
\pdfoutput=1

 \theoremstyle{mdpi}
 \newcounter{thm}
 \newcounter{ex}
 \newcounter{re}
\usepackage{microtype}
\usepackage{soul}

\Title{Baryon Number Transfer Could Delay Quark--Hadron Transition in
  Cosmology}

\Author{Silvio A. Bonometto $^{1,*}$ and Roberto Mainini $^{2}$}
\AuthorNames{Silvio A. Bonometto and Roberto Mainini}

\address{$^{1}$ \quad  The National Institute of Astrophysics (INAF)--- Astronomical Observatory of Trieste,
   \mbox{Departmnet of Physics},  University of Trieste, Via Tiepolo 11, Trieste 34143, Italy\\ 
  $^{2}$ \quad
  Department of Physics G.~Occhialini, Milano-Bicocca University, Piazza della
  Scienza 3, Milano 20126, Italy; mainini@mib.infn.it}

\corres{Correspondence: bonometto@oats.inaf.it}

\abstract{In the early Universe, {strongly interacting} matter was a
  quark--gluon plasma. Both lattice computations and heavy ion
  collision experiments, however, tell us that, in the absence of
  chemical potentials, no plasma survives at $T <\sim$$150\, $MeV. The
  cosmological {Quark--Hadron} transition, however, seems to have been
  a crossover; cosmological consequences envisaged when it was
  believed to be a~phase transition no longer hold. In this paper, we
  discuss whether even a crossover transition can leave an imprint
  that cosmological observations can seek or, vice versa, if there are
  questions cosmology {should address to} QCD specialists. {In
    particular, we argue }that it is still unclear how baryons (not
  hadrons) could form at the cosmological transition. A critical role
  should be played by diquark states, whose abundance in the early
  plasma needs to be accurately evaluated. We estimate that, if~the
  number of quarks belonging to a diquark state, at the beginning of
  the cosmological transition, is $<\sim 1:10^6$, its
  dynamics could be modified by the process of B-transfer from plasma
  to hadrons. In turn, by assuming B-transfer to cause just mild
  perturbations and, in particular, no entropy input, we study the
  deviations from the tracking regime, in the frame of SCDEW
  models. We find that, in~some cases, residual deviations could
  propagate down to primeval~nuclesynthesis.}


\begin{document}


\setcounter{section}{0} 
\section{Introduction}
The number of strongly interacting (s.i.) particles in the present
Universe appears negligible. They are mostly protons and neutrons,
sometimes embedded in nuclides, whose mutual distance, on~average,
exceeds 1$\, $m$\, $.  For the sake of comparison, in a cubic cm,
there are more than 400 photons.

The drastic decline of s.i. particle number density occurred at $T
\sim 150\, $MeV, when the primeval Quark--Gluon (QG) plasma ceased to
make part of the thermal soup. In the 1980s, when early lattice
results on QG plasma (see, e.g., \cite{Susskind 1979,Satz 1980}) seemed to
indicate that the transition from plasma to hadron gas was a real
first order phase transition, and a large number of papers was devoted to
study this transition in the cosmological context. A comprehensive
list can be found in the review paper \cite{Bonometto 1993}.

Attention then declined when lattice computations, including three quark
flavors and tentatively considering also dynamical fermions, begun to
make clear that, in the cosmological context, we had a crossover
transition. As a consequence, most of the envisaged observational
consequences on DM nature \cite{Witten 1984} and, namely, on
Big-Bang-Nucleosynthesis (BBN; see \cite{Bonometto 1985, Appegate 1985}), loosed their motivation.

We do not intend to revitalize these arguments here. The central point
of this paper will concern the $B$ transfer from primeval plasma to
hadronic gas, focusing on problems this could cause and formulating a
question that lattice computations, hopefully, can answer. In spite of
this, it is useful to recall why BBN was believed to be potentially
affected by the QH transition: the point is that BBN occurs when the
scale factor has increased less than a factor $10^3$ from the time of
the transition. If the transition had been first order, according to
Witten evaluations \cite{Witten 1984}, the baryon number $B$ could have tended
to remain in the plasma bubbles which finally shrank into {\it quark
  nuggets}, surviving until now and yielding DM. It soon became clear,
however, that even such remnants could not escape a final transition
into hadron matter; the process, however, yielded a strong $B$
concentration in the {\it nugget} sites. According to \cite{Bonometto 1985}, the
resulting $B$ inhomogeneities could yield proton peaks, lasting until
BBN, while residual neutrons reached homogenity by then.

  
In spite of the present common wisdom on the nature of QH transition,
the option of an {\it inhomogeneous} BBN has also often been
considered in recent literature, meaning that, at the opening of the
so-called {\it deuterium bottleneck}, proton distribution was still
inhomogeneous, so that neutrons had to flow back in proton
concentration sites before being able to yield $^2$H nuclides (see,
e.g., \cite{Keihanen 2002}).

Altogether, let us outline that there exist at least four contexts in
which the QH transition is debated. Besides the cited cosmological and
computational lattice contexts, there exists another astrophysical
possibility, the transition of neutron stars into quark stars
(e.g., \cite{Farhi 1984,Alcock 1986}), although no conclusive word has ever been said on
the energy gain allowed when $B$ is carried also by strange quarks,
besides $up$ and $down$ forming ordinary nucleons.
Finally, a highly significant new field of research has been opened up by
the study of heavy ion collisions, namely by LHC and RHIC experiments
(for a discussion see, e.g., \cite{Armesto 2006})

\begin{figure}[H]
\vskip -3.3truecm
\centering
\includegraphics[width=11.6cm]{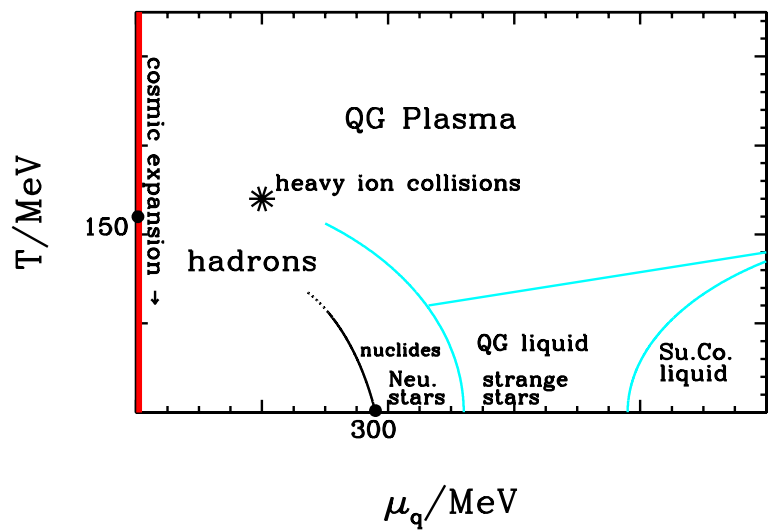}
\vskip -4.1truecm
\caption{A scheme of s.i. matter states and transition lines; the
  \textbf{red} vertical line at the extreme left shows the track of
  cosmological evolution, quite far from any other relevant area. We
  expect a phase transition to occur along the \textbf{black}
  curve. Any suggested phase transitions across other curves in
  \textbf{cyan} are hypothetical; the very existence of a QG liquid
  and of its superconducting phase is hypothetical~\cite{Alford 2009}.  $\mu_q$
  is the chemical potential associated to $B/3$, the baryon number in
  quarks.}
\label{diagram}
\end{figure}

Unfortunately, however, the only tool to relate laboratory and cosmic
data is numerical QCD, as~each cosmic context and laboratory outputs
are scarsely communicating; the very scheme in~Figure~\ref{diagram},
sometimes shown in general talks to show the connections among
different contexts, is still highly hypothetical. In turn, in spite of
the huge progress realized in lattice computation, they are still
unable to fully meet some simple datasets.

In spite of that, we can make two points that, in our opinion, did not
receive enough attention in previous literature, while being
potentially significant. In the next section, we shall deal with the
former of these points, concerning the transfer of $B$ from primeval
plasma to confined baryons; this will enable us to formulate specific
questions, concerning diquark states, that lattice computations and
heavy ion collision physics might try to answer. In Section \ref{ssr3}, we shall
somehow invert the perspective, and discuss the behavior of coupled
DM--DE, in SCDEW models (see, e.g., \cite{Bonometto 2012, Maccio 2015}), across the cosmological QH transition. The
basic merit and problem of these cosmologies is that they avoid
the conceptual problems of $\Lambda$CDM while keeping its successes,
so that they can be hardly discriminated from them. Accordingly, the
search of observables that SCDEW could affect bears a significance,
even though current observational errors still cover the
signals. Section \ref{ssr4} will be devoted to draw specific conclusions on
both topics.
\section{Baryon Number Transfer from Plasma to Baryons, and Likelihood
of Diquark States}

The {\it nuggets} of quarks envisaged by Witten, if real, would ease
the transfer of the extremely diluted baryon number from QG plasma
into three-quark baryons. If the QH transition is a crossover, {the
  transfer could be more intricate.

We can, however, imagine an ideal process, allowing a smooth
$B$-transfer. If the whole plasma could directly turn into hadron
gas, $\sim$$5\%$--$10\%$ of the newly formed gas would be made by
baryons and antibaryons, separately forming by sticking together three
suitably colored quarks and antiquarks. Successively, most baryons and
antibaryons mutually annihilate, thus leaving the fair amount of
residual baryons. Let us address this sequence of events as ``option
A''.

In turn, if only quark--antiquark bosonic doublets could form and no
(anti)baryonic triplets were synthetized, the transition would be
simply inhibited, as the Universe cannot remain with a residual of
unconfined quarks whose mutual distance is $\sim$1000 times the
confinement distance. Let us address this ideal scenario as ``option
0''.

What actually occurred is intermediate between these extreme options
and, unfortunately, closer to the ``option 0''. As is known, most of
the entropy of the QG plasma (initially $\sim$$69\%$ of the total)
has to turn into photon--lepton entropy, as the hadron gas will end up
with $<$$3\%$ of the total entropy (more~details on this latter
estimate are provided below). The point is then that the timing of
photon--lepton entropy production, far from being arbitrary, is
established by Friedmann equations, requiring the main reactions
occurring at the transition to be direct quark--antiquark
annihilations into photon--lepton pairs.

Freshly formed photons and leptons then infiltrate in the plasma.
They have no proper volume, however, and the presence of residual
quarks and antiquarks is no problem unless the overall (confined and
unconfined) quark number density shifts below $\sim$$T_c^3$. This
limit, however, cannot be overcome unless $B$ is fully transferred to
baryonic triplets. This must occur while the share of reactions
sticking together free (anti)quarks into (anti)baryonic triplets has
reduced by 1--2 orders of magnitude, so that we approach the ``option
0''.



The smooth crossover transition envisaged by lattice calculations and
tested in heavy ion collisions could then not coincide with what took
place in the cosmological context.

Of course, it is simply impossible that color charges become isolated.
If cosmic evolution approaches such a situation, what would occur is
that the color dielectric constant would reduce, so that the enlarged
color electric field produces lots of soft gluons. Because of the non-abelian nature of the gluon field, they will instantly form strings
connecting and neutralising the color charges. Their length being
large enough, they then decay in new quark--antiquark pairs leading to a
chain of hadrons. The cosmological meaning of this series of event is
that the Universe has gone out of equilibrium and the final output
would be a substantial entropy input.

$B$ transfer could be, however, facilitated by the presence of}
di--quark states. Truly confined quark couples arise only from the
union of quark and antiquark; however, besides these {\it straight}
couples, we ought to believe that, at least initially, there forms
some {\it homo} couples, made of two quarks or two antiquarks {and
  that a fair amount of them survives collisions and mutual
  annihilation}.

Similar anomalous couples, named {\it diquarks}, have indeed been
widely studied in recent literature. They were originally devised by
\cite{Ida 1966}
but really revitalized by \cite{Jaffe 2003},
when suggesting that they would provide
an explanation for an exotic baryon antidecuplet, the~$\Theta^+$,
whose evidence had been reported by the LEPS collaboration \cite{Nakano 2003};
objections were soon raised by \cite{Kabana 2004};
meanwhile, the matter has almost been settled, and, for a general
review, see, e.g., \cite{Moritsu 2016}).
Diquark evidence in lattice QCD was then soon outlined by \cite{Alexandrou 2002}.
For a general discussion on diquark properties, see \cite{Jaffe 2004,Wilczek 2004,Selem 2005}.

Let us add that more experimental evidence came in later years. For a
recent discussion, see \cite{Esposito 2015}
and references therein.
In turn, baryon form factor analysis also seems to indicate that
protons are to be seen as quark--diquark bound states (see, e.g.,
\cite{Segovia 2014}).
In~addition, diquarks are the central ingredient of the dense QG matter
yielding a color superconductor (see Figure~\ref{diagram}). 

{If quark--diquark collisions are the basic process responsible
  for $B$ transfer, we can}
perform a~rough estimate of which fraction $f$ of quarks should 
{keep} a~{\it homo} partner, when approaching $T_c$.  In~fact, the~
mean free path of
free quarks, for collisions with diquarks, reads
\begin{equation}
\lambda_2 \simeq {1 \over \sigma n_b f}    
\label{mfp}
\end{equation}
with $n_b \simeq 10^{-9} n$ (here, $n = (\zeta(3)/\pi^2)15.75\, T_c^3
\sim T^3_c$ is the quark number density at $T_c$) and $\sigma \sim
T_c^{-2}$. Accordingly, if we require $\lambda_2 \ll 10^3$cm ($\sim$$
1/10$ -- $1/100$ of the horizon at $T_c$), we conclude that,~if
\begin{equation}
f <\sim 10^{-6}~,
\label{fextim}
\end{equation}
baryon forming could significantly interfere with the dynamics of the
cosmological transition. Of~course, not all collisions will be
effective, for energetic and/or color matching reasons. This,~however,
does not change the order of magnitude of the estimate. Furthermore,
there will be the same rate of collisions with antiquark pairs, but
they hardly matter, as they are expected to yield just a straight
couple and leave a residual uncoupled $B$-carrying quark.

{Anyhow, unless all $B$ is transferred so that the ``risk'' of
  residual free quarks is over}, the overall {quark number}
density cannot decrease much below $T_c^3$.

Elementary processes being too slow to allow the Universe to remain in a
full equilibrium state are a recurrent feature in cosmic expansion.
The $^4He$ binding energy, e.g., is 28.3 MeV, but no helium can form
before $^2H$ nuclides, whose binding energy is 2.2 MeV, are massively
synthetized, so that they can undergo frequent collisions. An even
earlier event, when the Universe is expected to abandon equilibrium,
is inflation. At variance from the former example, among the effects
of inflation, there is a~huge input of entropy.

We can expect that confinement forces, even at the approaching of QH
transition, do not allow the average inter-quark distance to shift
substantially below $\sim$$T_c^{-1}$. This does not require all
s.i. matter to remain in the plasma phase. The requirement can be
fulfilled by the presence of hadronic particles, with a number density
$n_h \simeq T_c^{3}$.  At first sight, this might not seem like a strong
requirement: the number density of relativistic particles at a
temperature $T$ is indeed $ (\zeta(3)/\pi^2) \tilde g T^3$, and
$\tilde g=3$ for pions, while, close to $T_c$, we should also consider
the contribution of hadronic resonances.

However, neither pions nor resonances can be considered relativistic
at $T_c$. Furthermore, standard expressions refer to a gas of
point-like non--interacting particles. Mutual interaction impact was
modelized by \cite{Olive 1982},
while \cite{Karsch 1980} and \cite{Bonometto 1986}
outlined that particle {\it co}-volume could be even more impactant
as, although $\pi$s have no real {\it hard core}, they surely have a
physical size, $\sim$$T_c^{-1}$.
Altogether, in thermodynamical equilibrium, the number density of
$\pi$s, at~temperatures $\sim$$1.1$--$1.3 \times m_\pi$, is expected
to be one order of magnitude below the expression for point-like
uninteracting particles. Equations of state including a co-volume bear
a complex implicit form, the key point being that $p = f(n,T)$, so that
there can be different $p$-values at the same $T$, as~for ordinary
gases. One should then specify the process to detect equilibrium $n$-
and $p$-values for a given~$T$. 

If we forget $B$ conservation, or we suppose its transfer to occur
smoothly, entropy, previously carried by s.i. matter should massively
turn into lepton and photon entropy, rather than $\pi$s. Taking that
into account, we can devise two kinds of violations for such
equilibrium behaviour. (i)~{\it mild~violations}:~the~mutual distance
between hadrons keep $\sim$$T_c^{-1}$ as their number density is still
$T_c^3$, so that residual quarks carrying $B$, and not yet turned into
baryons, find a target to their confinement forces; (ii)~{\it strong
  violations}: if mutual distance between hadrons has to exceed
$T_c^{-1}$ not because of their turning into photon--leptons, but just
because of cosmic expansion, confinement forces could yield a~negative
pressure, thus causing a sort of mini-inflation able to keep a constant
s.i. matter density. At~variance from mild violations, this could
cause an entropy input, as though we had been facing am~out-of-equilibrium first order phase transition.

\vspace{8pt}
\noindent\emph{Bag-Like Models}
\vspace{8pt}

All that can be modeled, by using a suitable generalization of the old
MIT bag model, in a way coherent with lattice results. This
subsection will be devoted to discuss such modeling, which leads to
phenomenological expressions for pressure, energy and entropy
densities of s.i. matter. The upcoming expressions exhibit the nice
feature that they also allow us to model what might happen if $B$
transfer causes some delay in QH transition.

Let then $p = \Phi(T) T^4$, $\rho=E(T) T^4$, and $ \sigma = (p+\rho)/T
= \Sigma (T) T^3$, and remind the relation
\begin{equation}
E = 3\Phi+T\, d\Phi/dT,
\label{stateq}
\end{equation}
easily obtainable from the thermodynamical identity. Let us then
recall that 
\begin{equation}
g_{qg} = {7 \over 8} 2 \times 3 \times 3 + 2 \times 8 = 31.75
\label{gqg}
\end{equation}
is the statistical coefficient for quark and gluons (two spin states
for three colors and three flavors, for fermions, two~spin states for
eight bosons), and that the MIT bag model is soon obtained if we
assume~that
\begin{equation}
\Phi = {\pi^2 \over 90} g_{qg}[1-(\tilde T/T)^\alpha]
\label{phi}
\end{equation}
and that $\alpha = 4$ (the {\it bag} term then reads $ B =
(\pi^2/90)g_{qg}\tilde T^4$), but such an $\alpha$-value allows no agreement with
lattice outputs.


In order to compare expression (\ref{phi}) with lattice
data, let us first outline that it yields
\begin{equation}
E = {\pi^2 \over 30} g_{qg}[ 1 - (1-\alpha/3)(\tilde T/T)^\alpha]~,~~ \Sigma = 4
{\pi^2 \over 90} g_{qg} [1 - (1-\alpha/4)(\tilde T/T)^\alpha]~,
\label{Esigma}
\end{equation}
so that $E(T) $ is an increasing function only for $\alpha < 3$.

Much work has recently been done to obtain the s.i. matter state
equation close to $T_c$. For a recent review, see, e.g., \cite{Petreszky 2013}.
In Figure 1 of this paper, various outputs
of lattice computations are considered. In the sequel, we shall mostly
use ``$p4 (N_\tau=6)$'' outputs, due to these being those that extend to the
greatest value of $T$. In Figure \ref{power}, green triangles report
lattice outputs compared with the second expression (\ref{Esigma}) for
$\tilde T = 150\, $MeV and $\alpha = 1.6$. The fit can be improved if
more decimals are allowed, but, even so, the figure shows that
expression (\ref{Esigma}) is able to meet the 10 highest--$T$ points;
although purely phenomenological, such an expression surely performs
better than perturbative expression {or expressions tentatively
  including non perturbative effects by introducing suitable mass
  scales and/or making recourse to effective descriptions (see, e.g.,
\cite{Hietanen 2009,Laine 2006};
references related to different attempts can also be found in \cite{Borsanyi 2012}.
\begin{figure}[H]
\vskip -3.truecm
\centering
\includegraphics[width=9cm]{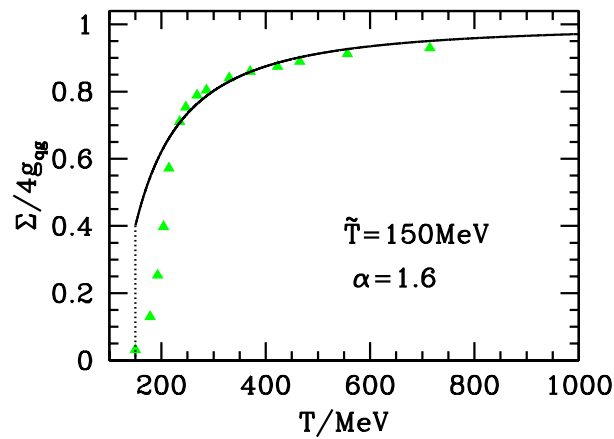}
\vskip -3truecm
\caption{Expression (\ref{Esigma}) vs. lattice outputs (see text)
  given by \textbf{green} triangles. The analytic curve does not meet the five
  lowest--$T$ points but is a fair approximation at greater
  temperatures. It is terminated at the temperature $\tilde T$, where
  $p$ would become negative.}
\label{power}
\end{figure}   

{Let us also add that a fit of the generalized bag--model} to the
``$p4(N_\tau=8)$'' leads to $\tilde T \simeq 160\, $MeV and $\alpha
\simeq 1.5$; results do not appear so impressive because the region of
the two highest--$T$ ``green'' points is not covered, but the
comparison between $\tilde T$ and $\alpha$ values used gives us an
idea of the actual uncertainty on fitting parameters. Let us add that
$asqtad$ outputs are not so far from $p4$ results, but concern a lower
$T$ range, where we cannot claim expression (\ref{Esigma}) to
apply. Altogether, expressions~(\ref{phi})~and~(\ref{Esigma}) appear
suitable to fit only higher $T$ points, although in a fairly wide
range. On the contrary, in the region of the five lowest-$T$ ``green''
points, expressions (\ref{phi}) and (\ref{Esigma}) are far from
lattice {\it data}. In~Figure \ref{power}, the plot of the analytical
curve is, however, terminated at $T=\tilde T$; below such a temperature,
expression (\ref{phi}) would return a negative pressure.

Let us now outline how we intend to interpret the fit to these {\it
  data}. The region where the fit nicely meets green points should
account for a gradual passage from QG plasma to a gas of confined
hadrons. The sixth green point region (from low $T$), below which the
fit no longer works so nicely, locates crossover completion: at lower
$T$, no {\it free quarks} remain; since then, confinement forces
act only inside hadrons. Accordingly, at smaller temperatures, the
density can decrease quickly, and fits could only be attempted,
perhaps, with expression based on hadron behavior, as those considered
by \cite{Karsch 1980}.

However, within such a context, the analytic curve does not totally
loose significance even in the low-$T$ region. It could be
interpreted as an upper limit to the entropy content in s.i. matter
form at such $T$, still allowing for a fluid where quarks (confined or
unconfined), however, lay at distances $<\sim$$T_c^{-1}$, so~that gluons
arising from {\it residual free quarks} can still be reabsorbed ``in
time''. This would be the region of {\it mild violations} of the
equilibrium entropy sharing among cosmic components, possibly required
by the persistence of $B$ carrying free quarks. On the contrary, once
attaining a negative pressure, we~approach a regime where a sort of
mini-inflation would occur, with creation of quark--antiquark pairs
by confinement forces, so that the overall quark number density does
not shift below a limiting value. This would be the region of {\it
  strong violations}.

Of course, all of this is just a tentative modeling for situations
still out from full numerical control. However, it lends us a tool to
model what happen in the case of delayed baryon formation, according
to the value of the abovementioned parameter $f$.

We believe that suitable lattice computations can provide $f$-values
with a fair approximation. Being still unavailable, we shall consider
here the {\it reasonably worst} option: that cosmic evolution occurs
along the analytical curve until $p=0$. By then, the $B$ transfer
is completed, so that no mini-inflation and entropy input occur,
while the following expansion takes place with a rapid transformation
of s.i. matter entropy into photon--lepton entropy. For what concerns
cosmic evolution, this is similar to what was found to occur in a
first order phase transition occurring exactly at $T_c$ and,
therefore, without~entropy input.  The above expressions (\ref{phi})
and (\ref{Esigma}) will then enable us to integrate cosmic expansion
equations during such an epoch.

In the next section, we shall reconsider such problems in the frame of
SCDEW cosmologies.

\section{An Analytical Description of the Cosmological QH Transition,
and Coupled DM-DE Equilibrium Recovery in SCDEW Models}\label{ssr3}

Wishing now to consider the transition within a cosmological context,
let us recall the cosmic background metric if (almost) flat, even
today, while the density parameter of matter (baryons~plus~DM) is
$\Omega_m \simeq 0.3$; detailed values can be found, e.g., in the
recent report of the Planck collaboration \cite{Planck 2015}. 
This~suggests the existence of a component or phenomenon, denominated
as DE, filling the gap. To~be~also consistent with SNIa data (see, e.g.,
\cite{Kowalski 2008}),
it should also cause cosmic expansion to~accelerate.

The $\Lambda$CDM model, initially revived to meet acceleration data,
has then become a sort of {\it standard cosmology}, not only because
it fulfills the above requirements, but also for meeting much more
data well beyond cosmological acceleration and background
composition, which can be suitably accommodated inside it. However,
such a model, assuming that DE has a state equation $p = w\, \rho$ with
$w \equiv -1$ ($p,~\rho$:~average cosmic pressure and energy density),
implies a mess of paradoxes and conundrums, so~that it is mostly
considered just a sort of {\it effective} model, hiding a deeper
physical reality that present data are insufficiently detailed to
discriminate.

While new experiments are running to enrich the datasets (see, e.g.,
LSST (http://www.lsst.otg/lsst/)
and {\sc Euclid} \cite{Laureijs 2011}, 
much
work has been devoted to forge cosmological models, overcoming
$\Lambda$CDM conundrums while~being indistinguishable from it, within
the context of available datasets.

\subsection{SCDEW Cosmologies}
Cosmologies involving Strongly Coupled DM and DE, plus warm DM
(SCDEW), are a possible option among them. They start from the
discovery of an attractor solution for a Friedman equation in the early
radiation-dominated Universe, which, at variance from any other
models, involves early non--radiative components. Let us summarise how
this works.

The energy density of non-relativistic uninteracting DM and a purely
kinetic scalar field $\Phi$, in a radiatively expanding environment,
dilute as $a^{-3}$ and $a^{-6}$, respectively. Here, $a$ is the scale
factor and
\begin{equation}
ds^2 = a^2(\tau) (d\tau^2 - d\lambda^2)
\label{metric}
\end{equation}
is the background metric, $\tau$ and $d\lambda$ being the conformal
time and the comoving distance element, respectively. However, if DM
and $\Phi$ are coupled and energy flows from DM to $\Phi$, both of the above
components could be led to dilute as $a^{-4}$, just as background
radiation.

This expectation is confirmed by precise calculations, showing that
the Lagrangian coupling needed to obtain this result bears a
Yukawa-like form
\begin{equation}
{\cal L}_I = - g\, m_p \exp(-b\Phi/m_p) \bar \psi \psi~.
\label{yukawa}
\end{equation}

Here: $\psi$ is a spinor field yielding DM; $m_p$ is the Planck mass;
and $g$ and $b$ are suitable constants, the former one being subject to
loose constraints, while the latter one can be used to define
the {\it coupling~constant}
\begin{equation}
\beta = b \sqrt{3/16 \pi}~.
\label{beta}
\end{equation}

It can then be shown that, if DM and $\Phi$ have density parameters
\begin{equation}
\Omega_c = 1/2\beta^2 ~~~{\rm and} ~~~ \Omega_\Phi = 1/4\beta^2~,
\label{omgas}
\end{equation}
the cosmic expansion proceeds along an attractor, i.e., both radiation
and the coupled $DM$-$\Phi$ component keep stable proportions, all
diluting as $a^{-4}$. Notice that all this implies that $\beta^2 >
3/4$, and this is why this approach implies a {\it strong coupling}
between DM and $\Phi$. Of course, if $\beta \equiv \sqrt{3}/2 =
0.8660$, there is no room for ordinary radiation. At the passage
through big--bang nucelosynthesis (BBN) and at the last scattering
band (LSB), the coupled $DM$--$\Phi$ component is a sort of {\it dark
  radiation} (DR). In~Figure~\ref{DR}, we plot the values of $\beta$
corresponding to a given amount of DR, expressed in terms of extra
neutrino~species.
\begin{figure}[H]
\vskip -3.truecm
\centering
\includegraphics[width=9cm]{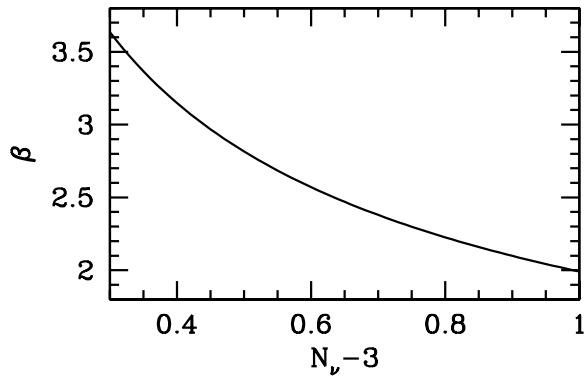}
\vskip -3.4truecm
\caption{Values of the coupling $\beta$ corresponding to given values
  of {\it Dark Radiation}, expressed in terms of extra neutrino
  species. This plot is valied during BBN.}
\label{DR}
\end{figure}   

BBN is consistent with data even if DR accounts for up to about one
extra neutrino species.  Accordingly, $\beta $ values as low as two
appear acceptable. A stronger constraint comes from CMB data. As shown
by \cite{Bonometto 2012}, 
the actual amount of DR has (slightly) increased, at the passage through the LSB, because of the
end of the self-similar expansion regime after matter--radiation
equivalence. At the LSB, one extra neutrino species corresponds then to
$\beta=2.5$. According to \cite{Planck 2015}, 
at the 95$\, \%$ confidence level, the number of extra neutrino species
allowed are~$\simeq$$0.7$.  Taking one extra neutrino species exceeds
this limit. However, besides being clearly consistent with data
within three $\sigma$s, such a greater value of $N_\nu$ would favor
Hubble constant values $\sim$$0.72$, consistent with direct
observations. On the contrary, the best fit of the Planck
collaboration data disagrees with them. Therefore, we can set the
lower limit to coupling at $\beta = 2.5~.$

Here below, most results will be shown for $\beta=8$. However, when
considering the impact of QH transition on BBN, we shall compare
results for $\beta = 2.5$ and 10.

Altogether, the dynamical equations derived by the coupling Lagrangian
$\cal L$$_I$ and the kinetic terms can be set in the form

$$
\dot a = (8\pi/3)^{1/2} (\rho_c + \rho_\Phi + \rho_r)^{1/2} a/m_p~,
$$$$
\dot \rho_c = -3\rho_c/\tau - \rho_c \Phi_1 b/m_p~,
$$
\begin{equation}
\dot \Phi_1 = -2\Phi_1/\tau + \rho_c a^2 b/m_p~,
\label{eqs}
\end{equation}
provided we assume DM to be fully non--relativistic, its density being
$\rho_c$. Here, $ \Phi_1 = \dot \Phi$, dots yielding differentiation in
respect to $\tau$, while $\rho_r$ is the density of the whole {\it
  thermal soup}. As the decreasing cosmic temperature shifts below the
mass value of any particles, the standard event is that they
annihilate by keeping a constant value of cosmic entropy
\begin{equation}
S = {4 \over 3} {\pi^2 \over 30} g\, T^3a^3,
\label{entropy}
\end{equation}
even though
\begin{equation}
g = {\cal N}_b + {7 \over 8}{\cal N}_f
\label{g}
\end{equation}
(${\cal N}_{b,f}:$ number of independent spin states of bosons,
fermions with $m < T$) has decreased.  As $S$ is constant, and $\rho =
(\pi^2/30) gT^4,$ it shall be
\begin{equation}
\rho \propto  g^{-1/3} a^{-4},
\end{equation}
so that ``radiation'' deviates from a pure $a^{-4}$ diluting by
exhibiting an upward jump due to $g$ decrease.

When this happens, $\rho_c$ and $\rho_\Phi$ also should shift upward,
so that $\Omega_c$ and $\Omega_\Phi$ preserve the values~(\ref{omgas}). In other terms, the attractor solution, perturbed by
$g$ decrease, must rearrange; but the process may take some time and
its duration can be tested just by inserting $\rho_r$ behavior in
Equation~(\ref{eqs}).

The greatest jump the attractor may then have to face is at the QH
transition, when $g$ shall exhibit a drastic decay within quite a
short temperature interval. An even more drastic event will occur if
the {\it reasonably worst} option, considered in the previous section, is
true. In this case, the decrease of the effective $g$ shall take place
while $T$ stays constant at $\tilde T$ along the dashed line in
Figure \ref{power}.

Before studying the behavior of densities across such a transition, let
us briefly complete the reminder on SCDEW cosmologies. The attractor
solution for Equation~(\ref{eqs}) was found by Bonometto~et~al. (2011),
who also studied the behavior of cosmic components at lower $T$s,
showing that it easily meets the observed proportions of background
components. Of course, this requires that the $\Phi$ field shifts from
kinetic to potential at a suitable time. This can be obtained without
making recourse to a specific (tracker) potential, as detailed in the
above paper. Bonometto and Mainini~\cite{Bonometto 2014} then studied the evolution of
density fluctuations within such cosmologies, finding that CMB
anisotropies and polarization are substantially indistinguishable from
$\Lambda$CDM and calculating the linear transfer function. More
recently, Macci\'o et al.~\cite{Maccio 2015} performed $N$-body simulations for
some model versions, finding that it reproduces $\Lambda$CDM findings
at scales above average galaxy sizes, while substantially easing long
standing $\Lambda$CDM problems such as dwarf galaxy profiles, MW
satellites, and galaxy concentration~distribution.

Altogether, all nice $\Lambda$CDM outputs are practically overlapped
by SCDEW, which, however, allows the abovementioned improvements over
lower scales while eradicating all conceptual problems going with a
simple-minded $\Lambda$CDM scheme.  A number of subtle tests can, however, be
devised, enabling us to distinguish between SCDEW and
$\Lambda$CDM. As expected, they shall be possible only when a much
higher level of precision will be attained by observations. It is also
fair to outline that SCDEW involves a number of extra parameters, but
no fine tuning is needed for any of them.  A final point is that
the Lagrangian $\cal L$$_I$ could also account for post-inflationary
reheating and the field $\Phi$, besides of beind today’s DE, could
also play an important role in the inflationary process.

\subsection{Attractor Behavior at the Cosmological QH Transition}
Taking into account expressions (4) and (6), we find that, in the
proximity of the QH transition, the $\rho_r$ term at the r.h.s.
of the first Equation~(\ref{eqs}) shall read
$$
\rho_r = {\pi^2 \over 30} \left\{g_{\gamma,lep} +
g_{ql}\left[1-\left(1-{\alpha \over 3} \right) \left(\tilde T \over T
    \right)^\alpha \right] \right\} T^4 ~~~~~~~ ({\rm for} ~ T > \tilde T
~{\rm and}~~ a< a_1),~
$$$$
\rho_r = {\pi^2 \over 90} \left[ \left(4 g_{\gamma,lep} + \alpha
  g_{ql} \right) \left(a_1 \over a \right)^3 - g_{\gamma,lep} \right]
\tilde T^4
~~~~~ ({\rm for} ~ T \equiv \tilde T
~{\rm and}~~ a_1< a < a_2),
$$
\begin{equation}
~~
\rho_r = {\pi^2 \over 90} g_{\gamma,lep} T^4
~~~~~~~~~~~~~~~~~~~~~~~~~~~~~~~~~~~~~~~~~~~~~~~~~~~~~~~~~~~~~~~~~
({\rm for} ~ T < \tilde T ~{\rm and}~~ a> a_2).
\label{rhos}
\end{equation}

Here,~$g_{\gamma,lep}$ is the statistical weight of lepton--photon spin
states. Amongst leptons, we also include $\mu$ particles; their
progressive annihilation, when $T < m_\mu$, will be disregarded, in
order not to confuse its effects of QH transition.

Owing to $S$ conservation before, during, and after the transition,
the scale factor at its end shall~be
\begin{equation}
a_2 = {T_0 \over \tilde T} a_0 \times \left(g_0 \over g_{\gamma,lep}
\right)^{1/3}~,~~~ {\rm with}~~ g_0 = 2(1+2.625/2.75^{4/3}),
\label{a2}
\end{equation}
accounting for the effective number of spin states at the present
cosmic temperature $T_0$ (scale factor~$a_0$) by assuming three ($\sim$massless) neutrino flavors. This scale factor is reached after a
temperature {\it plateau} (at~$\tilde T$), starting when the scale factor
is
\begin{equation}
\label{a1}
a_1 = a_2 \left( g_{\gamma,lep} \over g_{\gamma,lep} + g_{qg}/4 \right)^{1/3}~.
\end{equation}

Of course, as soon as $(\tilde T/T)^\alpha$ acquires a significant
value (in comparison with unity), the standard cosmological behavior
$aT = {\rm const.}$ is violated. In Figure \ref{Ta}, we show the size of
such a violation for $\alpha=1.6$ and compare it with the case $\alpha = 2$.
This behavior is obtainable just from $S$ conservation and is only
mildly dependent on $\beta$, provided it is not too close to its lower
limit $\sqrt{3}/2$. The plot was obtained, however, for $\beta = 8,
$ {as all plots here below}.
\begin{figure}[H]
\vskip -2.truecm
\centering
\includegraphics[width=9cm]{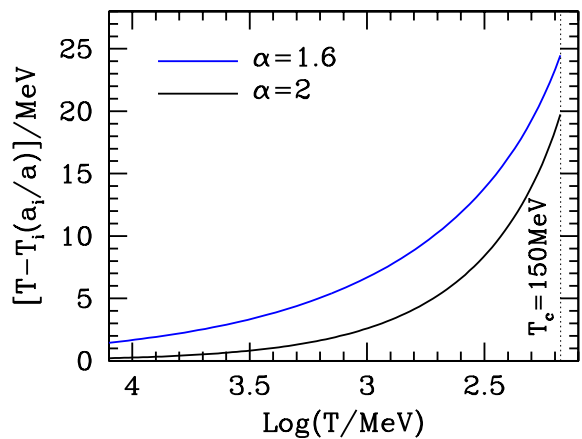}
\vskip -3.1truecm
\caption{Violation of the condition $aT = {\rm const.}$ at the
  approaching of QH transition. $T_i$ ($a_i$) is any large temperature
  (small scale factor) where asymptotic freedom still holds.}
\label{Ta}
\end{figure}   

In particular, in Figure \ref{aTvsa}, the same effect is seen in more
detail all through the transition. When~hadrons become substantially
negligible, $aT$ recovers a fully constant behavior vs. $a$. 
\begin{figure}[H]
\vskip -1.5truecm
\centering
\includegraphics[width=8cm]{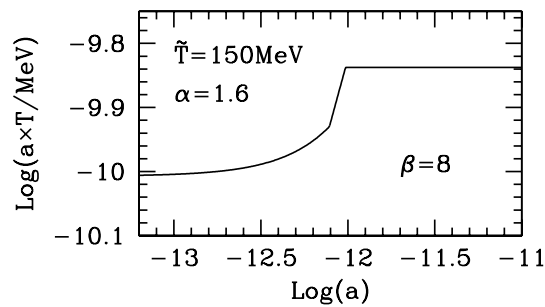}
\vskip -5.truecm
\caption{The product $aT$ is plotted vs. $a$ all through the
  transition. The angular points are an effect of the approximations
  used to model the transition and, in particular, of the assumption
  that quark carried $B$ suddenly ceases to exist all through each
  horizon.}
\label{aTvsa}
\end{figure}   

During a radiative expansion, when $\rho a^4 = {\rm const.}$, the
Friedman eq. reads
\begin{equation}
{1 \over \tau^2} = {8 \pi \over 3 m_p^2} \rho a^2 \propto {1 \over a^{2}} ~,
\end{equation}
so that $a \propto \tau$. When $\rho a^4$ is no longer constant, as
during the QH transition, this proportionality can be violated.  In
Figure \ref{aotauvsa}, we show such violations through QH transition.
\begin{figure}[H]
\vskip -2truecm
\centering
\includegraphics[width=8cm]{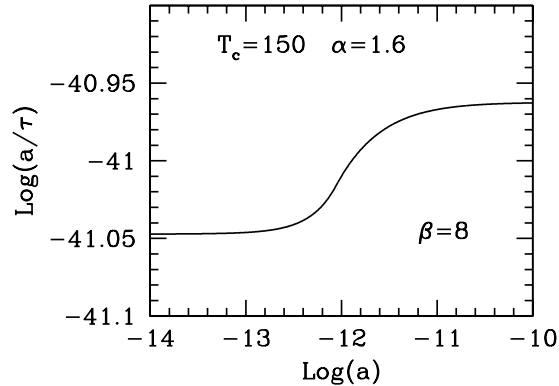}
\vskip -3.truecm
\caption{Scale dependence of the $a/\tau$ ratio during the QH transition. }
\label{aotauvsa}
\end{figure}   

\subsection{Density Anomalies Caused by QH Transition}
Let us finally discuss the effects of QH transition on early
$\Omega_c$ and $\Omega_\Phi$ values. QH transition is likely to be the
greatest push out of the attractor solution that cosmic components
receive during early expansion. It is shown in Figure \ref{rhos} both
for $\beta = 2.5$ and $\beta = 10\, .$

The initial effect of the transition is to decrease the ratio
$(\rho_c+\rho_\Phi)/\rho_r$ by $\sim$$12\%$, almost independently of
$\beta$, showing that the increase of radiation density, at QH
transition, is only partially followed by coupled DM and $\Phi$
densities. The subsequent rebounce upward, however, is wider, its~amount being slightly greater for smaller $\beta$ values, and occurs
when the temperature has already lowered to $T$ $\sim$ 20 MeV.

Residual perturbations at the eve of BBN depend still more on $\beta$
and can be better appreciated in Figure \ref{rhos1}. Just at the time
of neutrino decoupling ($\sim$900 keV), we have a second
minimum. The fraction of cosmic materials in the radiative component
is then almost steadily increasing until the reaching of deuterium
bottleneck.

In general, the possible impact on BBN is obviously greater for
lower $\beta$. However, even for $\beta=2.5$, the coupled DM--$\Phi$
component has a limited impact on the overall density, approximately
equivalent to a half extra neutrino species, seemingly consistent with
primeval abundance limits. On~top of that, we have residual
$\rho_r/\rho_{total}$ oscillations, so that the scale dependence of
both overall and radiation expansions exhibit incoherent discrepancies
in respect to a purely radiative regime. For $\beta = 2.5$,
discrepancies can be $\cal O$$(0.5\, \%)$ and almost steadily continue
all through the period going from $\nu$-decoupling to the synthesis
of $^2H$ and other nuclides. Quantitatively, such deviation is not
much less than half of what is due to electron--positron annihilation.

\begin{figure}[H]
\vskip -1.truecm
\centering
\includegraphics[width=7.6cm]{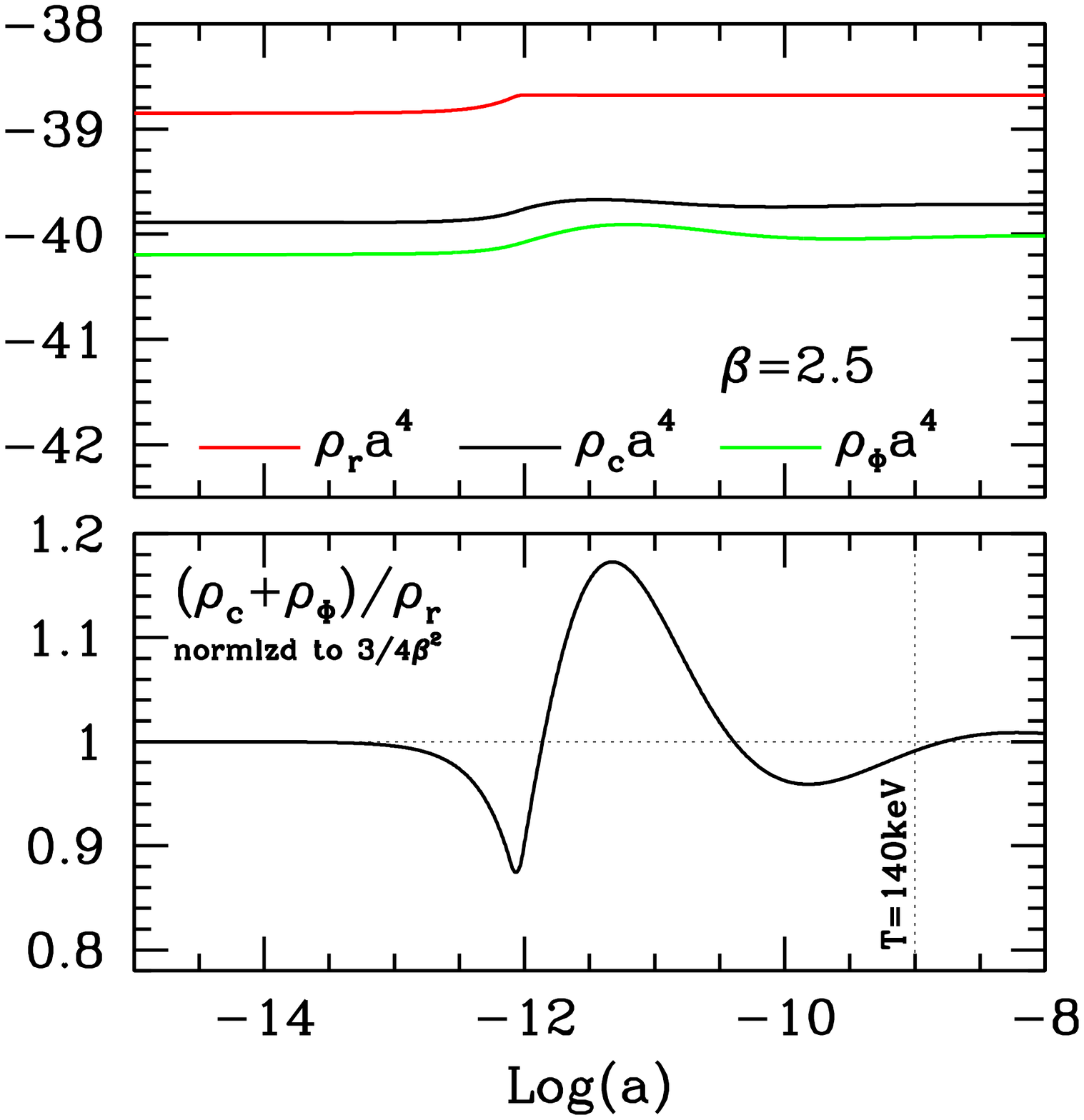}
\includegraphics[width=7.6cm]{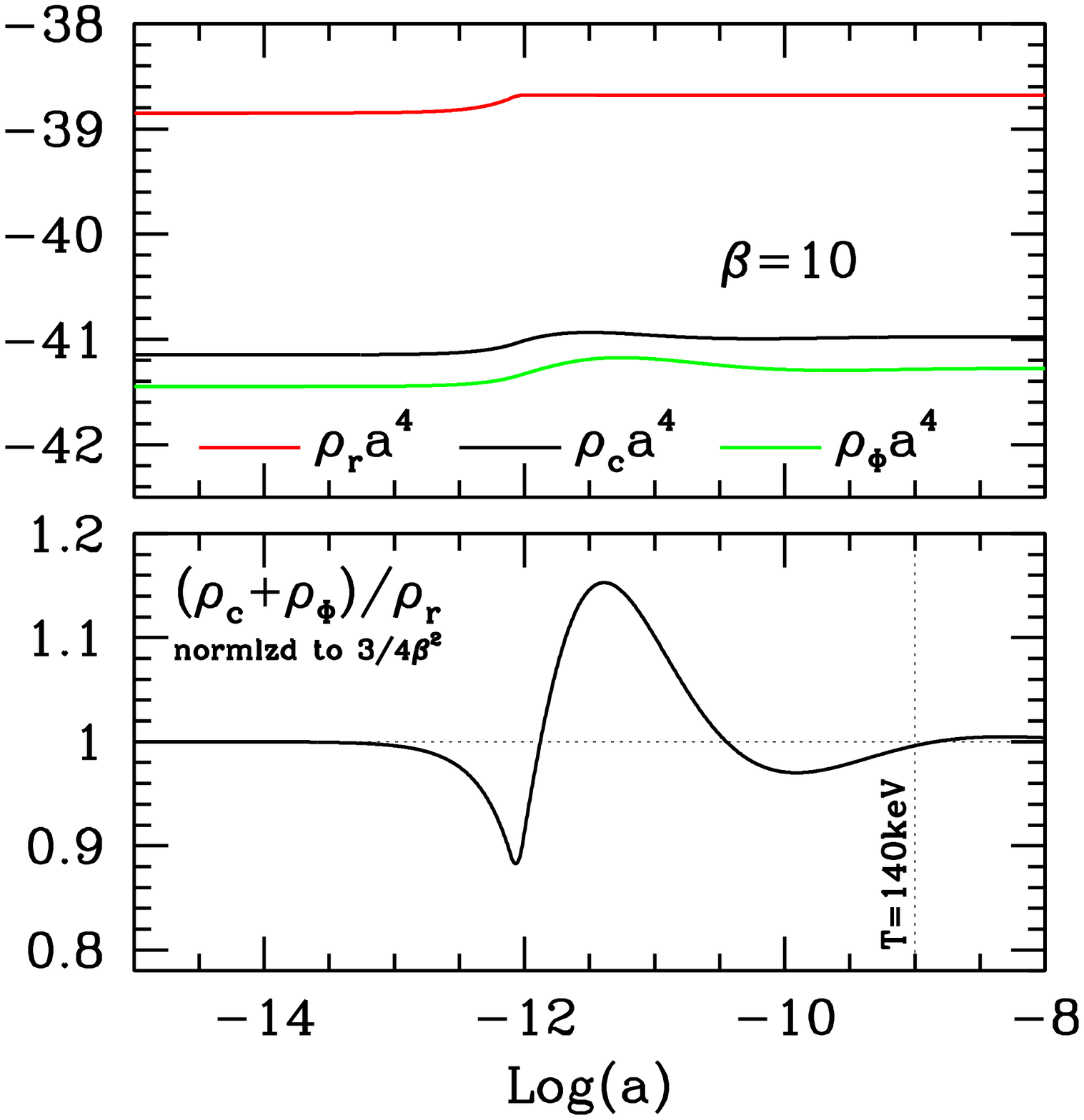}
\vskip -1.5truecm
\caption{QH transition causes oscillations of the early density
  parameters $\Omega_c = 2\Omega_\Phi = 1/2\beta^2$.  The~system of
  Equation (\ref{eqs}), when $\rho_r$ is given by expression
  (\ref{rhos}), yields the results shown here for $\beta = 2.5$ and
  $10$. In the upper frames, the logarithms of $\rho_{r,c,\Phi}a^4$
  are plotted. Their deviation from the constant is also outlined in
  the lower frames. The normalized ratio $(\rho_c+\rho_\Phi)/\rho_r$
  exhibits a mild $\beta$ dependence, in the initial oscillations,
  with maximum deviation by $\sim$$+15$\%$,+18$$\%$, when $T \simeq
  20\, $MeV,
(if~$T_c \simeq 150\, $MeV). For the behaviors close to BBN, see the
  next figure.}
\label{rhos}
\end{figure}   
\unskip
\begin{figure}[H]
\vskip -2.truecm
\centering
\includegraphics[width=8cm]{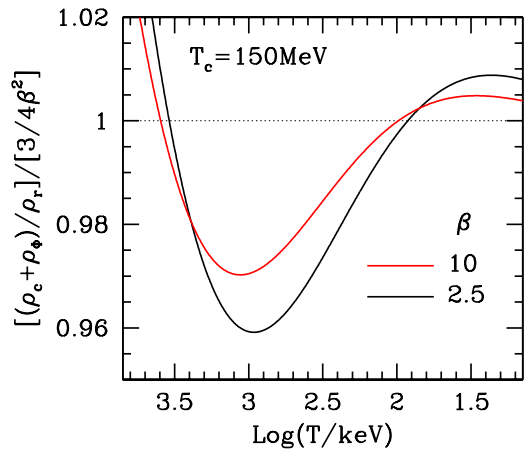}
\vskip -1.5truecm
\caption{Magnification of the lower plots in previous figures for the
  region of BBN, with $T$ (instead~of~$a$) as ascissa. The~second
  minimum, as also shown 
{in the previous Figure}, occurs at $T \sim 900\,
  $keV, quite close to neutrino decoupling. The scale dependence of
  the radiative component between neutrino decoupling and the region
  of early $^2H$ synthesis undergoes a further deviation from
  radiative expansion, up to $\cal O$$(0.5\%)$, due
  to electron--positron annihilation. Such deviation depends on
  $\beta$ being stronger when the coupled DM-$\Phi$ component is
  denser.}
\label{rhos1}
\end{figure}   

Of course, no major changes in BBN predictions can be expected because
of these density shifts. The point is that, knowningly, this kind of
deviation from the standard model were never tested in BBN numerical
experiments.  Even though such variations are mild, however, they
would
be a signature of SCDEW models in respect to $\Lambda$CDM. 

In general, discrepancies between SCDEW and $\Lambda$CDM results
are mild. This is a SCDEW peculiarity and a major merit. In fact, no other
variations within the frame of available data is~appreciable. A final word on variations in the estimated light nuclide
abundances, arising from such slight deviations from a
$\Lambda$CDM-like expansion regime, can hardly be said without direct
inspection.  It~is not unreasonable to expect that this effect is
also rather small, but this is the kind of discrepancy that may enable
us to say a final word on SCDEW being the true underlying cosmology
yielding an apparent $\Lambda$CDM.

\section{Discussion}\label{ssr4}
The Universe arrives at the QH transition with a tiny excess of
quarks, in respect to anti-quarks. On~average, then, {in any
  volume with $\sim$1000 (anti)quarks across, there is one extra quark
  carrying~$B$.}
{As the direct formation of quark triplets is a rare process, when
  most quark--antiquark pairs undergo mutual annihilation into
  photon--letpon pairs, in} order to transfer $B$ from the QG plasma
to the hadronic gas, we 
{should probably} rely on the formation of diquark couples, aside
from the straight quark--antiquark proto-hadrons, when the transition
has a start; i.e., statistical equilibrium should prescribe that a~sufficient fraction of quarks enter homo rather than hetero couples at
the {beginning} of the transition. Then, baryons can form through
ordinary two-body collisions {between quark and diquark},
provided that the center-of-mass energy is sufficiently high and there
is correct color matching. We estimated that the minimal requirement
to allow this to happen is that the fraction $f$ of quarks embedded in
diquarks at the {beginning} of the QH transition is not less than
{$1:10^6$}.

This, however,
pushes us to consider the
possibility that, until $B$ has not fully
transferred from QG plasma to hadron gas, s.i. materials could not
dilute to densities significantly $<$$T_c^3$, in order to maintain mutual
inter-quark distances $\sim$$T_c^{-1}$. This requirement could cause a
deviation of s.i. matter and entropy densities ($\rho_{qg}$ and
$\sigma_{qg}$) from straightforward lattice predictions.

Lattice computation shows that the inter-quark distance gradually
increases at the approaching of a suitable temperature $T_c$, in order to
then undergo a faster decrease about such $T_c$. We showed that
the~expressions
\begin{equation}
\rho_{qg} = {\pi^2 \over 30} g_{qg}\left[ 1 - (1-{\alpha \over
    3})\left(\tilde T \over T \right)^\alpha \right] T^4~,~~
\sigma_{qg} = 4 {\pi^2 \over 90} g_{qg} \left[1 - (1-{\alpha \over
    4})\left( \tilde T \over T \right)^\alpha \right] T^3~,
\label{Esigma1}
\end{equation}
for energy and entropy density, provide a fair fit to higher--$T$
lattice outputs with $\alpha = 1.5$--1.6 and $\tilde T \sim 150\,
$MeV. These expressions are a generalization of the old MIT bag model,
which is obtained again if $\alpha=4$. They, however, cease to meet lattice
results in the fast decreasing regions. In turn, the related pressure
expression
\begin{equation}
p_{qg} = {\pi^2 \over 90} g_{qg}\left[1-\left( \tilde T \over T
  \right)^\alpha \right] T^4,
\label{phi1}
\end{equation}
yields a faster decrease of pressure $p$, implying a {\it negative}
pressure at $T < \tilde T$. 

It should be outlined that this somehow strengthens the significance
of expressions (\ref{Esigma1}) and~(\ref{phi1}). In the
cosmological context, a negative pressure could lead to a sort of
mini-inflation---s.i. matter being produced by confinement forces,
rather than allowing for a plasma with mutual particle distances~$\gg$$
T_c^{-1}$. However, even without considering such an extreme option in the
temperature interval between $T_c$ and $\tilde T$, these expressions
are still useful to model equilibrium deviations if and when
$B$-transfer is delayed. This is what we did in this work in order to
estimate the cosmological effects of a~possible diquark fraction $f <
10^{-6}$.

In more detail, we considered a specific example of the transition
being delayed just until $\tilde T$ is reached, considering it as the
{\it worst reasonable} option. Then, in the frame of SCDEW models, the
delay yields significant oscillations in the ratio between radiative
materials and the coupled DM-$\Phi$ component. We provide a numerical
solution of the three-equation system yielding these oscillations,
showing that they may propagate down to BBN.

We then showed that we can expect a (mild) deviation from a purely
radiative expansion between neutrino annihilation and $^2H$
synthesis. Quantitatively, it can reach half of the deviation due to
electron--positron annihilation. However, to our knowledge, such a kind
of deviation from standard $\Lambda$CDM models was never numerically
tested.

Readers should be reminded, however, that, at present, no available data
allow us to discriminate between SCDEW and $\Lambda$CDM, the main
merit of SCDEW being its capacity to faithfully reproduce $\Lambda$CDM
data by avoiding its conceptual problems while also relating the
inflationary field with DE (as a matter of fact, SCDEW also eases some
$\Lambda$CDM problems below the galactic scale; Macci\'o et al. \cite{Maccio 2015}).

It is therefore important to devise the areas where more refined data
could allow us to finally discriminate between these two sets of
cosmologies. In this paper, we outlined that QH transition could be
entangled with finding one such area.

\section{Conclusions}

In summary, we 
are led to
ask QCD specialists and BBN cosmologists two separate questions:
(i)~is~it possible to 
provide a really reliable
estimate of the fraction of quarks
{in} diquark couples {along} the crossover transition?  (ii){electron--positron annihilation}
{causes} a deviation from radiative matter dilution $\propto a^{-4}$
during BBN.
How much smaller can a further deviation be during the same period in
order to be appreciable through observational light nuclide
abundances? As a matter of fact, we can expect that the QH transition
causes an echo extending down to BBN, with significant deviations from
radiative matter dilution~$\propto a^{-4}$, {ranging between 1/4 and
  1/2} of what is due to electron--positron annihilation.

\vspace{6pt} \acknowledgments{The authors would like to thank Matteo
  Viel, Claudio Destri and Michele Pepe for discussions.}
\authorcontributions{The contributions by each one of the two authors
  could be hardly disentangled.}
\conflictofinterests{{The authors declare no conflict of interest.}}


%
%



\bibliographystyle{mdpi}
\renewcommand\bibname{References}

\bibliographystyle{mdpi}

\end{document}